\documentclass[useAMS,usenatbib]{mn2e}
\usepackage{aas_macros}
\usepackage{amsmath, amssymb}
\usepackage{graphicx}
\usepackage[caption=false,lofdepth,lotdepth]{subfig}
\usepackage{url}

\begin{document}

\title[Luminous X-rays from SN \& CSM shell interactions]{Super-luminous X-ray Emission from the Interaction of Supernova Ejecta with Dense Circumstellar Shells}
\author[T. Pan, D. Patnaude, A. Loeb]{Tony Pan$^1$, Daniel Patnaude$^1$, Abraham Loeb$^1$\\
$^1$Harvard-Smithsonian Center for Astrophysics, 60 Garden Street, Cambridge, MA 02138, USA\\
}

\pagerange{\pageref{firstpage}--\pageref{lastpage}} \pubyear{2013}
\maketitle
\label{firstpage}

\begin{abstract}

For supernova powered by the conversion of kinetic energy into radiation due to the interactions of the ejecta with a dense circumstellar shell, we show that there could be X-ray analogues of optically super-luminous SNe with comparable luminosities and energetics.  We consider X-ray emission from the forward shock of SNe ejecta colliding into an optically-thin CSM shell, derive simple expressions for the X-ray luminosity as a function of the circumstellar shell characteristics, and discuss the different regimes in which the shock will be radiative or adiabatic, and whether the emission will be dominated by free-free radiation or line-cooling.  We find that even with normal supernova explosion energies of $10^{51}$ erg, there exists CSM shell configurations that can liberate a large fraction of the explosion energy in X-rays, producing unabsorbed X-ray luminosities approaching $10^{44}$ erg s$^{-1}$ events lasting a few months, or even $10^{45}$ erg s$^{-1}$ flashes lasting days.   Although the large column density of the circumstellar shell can absorb most of the flux from the initial shock, the most luminous events produce hard X-rays that are less susceptible to photoelectric absorption, and can counteract such losses by completely ionizing the intervening material.  Regardless, once the shock traverses the entire circumstellar shell, the full luminosity could be available to observers.

\end{abstract}

\label{lastpage}


\section{INTRODUCTION}

An interesting question is whether there could be X-ray counterparts to super-luminous supernova, with comparable luminosities and/or total energy emitted.  Excluding the energy emitted by neutrinos, most core collapse supernova (SN) have explosion energies of order $10^{51}$ ergs, but usually only $10^{49}$ ergs of that energy is released as optical radiation during the supernova, with typical peak luminosities not exceeding $\sim 10^{43}$ erg s$^{-1}$.  However, numerous super-luminous supernovae with luminosities $\gtrsim 10^{44}$ erg s$^{-1}$ were discovered over the past decade \citep{Gal-Yam2012}, some of which had total radiated energies $\sim 10^{51}$ ergs, e.g. SN 2003ma \citep{Rest2011} and SN 2006tf \citep{Smith2008}.  Although a few of these events may be powered via radioactive decay, e.g. SN 2007bi \citep{Gal-Yam2009}, a distinct majority of super-luminous supernova require some other mechanism to power their radiative output.   

One of the main mechanisms\footnote{The other main mechanism is the outward diffusion of deposited shock energy in optically thick CSM, i.e. the shock breakout, which can also produce X-rays; see Section \ref{SectionDiscussion}.} invoked to convert a larger fraction of the large explosion energies into optical emission is via the strong interaction between the expanding supernova ejecta and massive circumstellar material (CSM) previously expelled by the star \citep{Smith2007}.  Similarly to Type IIn supernova, the bulk kinetic energy of the ejecta is converted back into radiation via strong shocks \citep{Chevalier1994}.  The energetics of this process can be understood via the following toy model: if two objects of mass $M_a$, $M_b$ with velocities $v_a$, $v_b$ collide and stick together, conservation of energy and momentum dictates that the kinetic energy lost from the inelastic collision will be:
\begin{equation}
\Delta E_{kinetic} = \frac{1}{2} \frac{M_a M_b}{M_a + M_b} \left(v_a - v_b \right)^2
\end{equation}
If $v_a \gg v_b$, and the lost kinetic energy is converted to radiation with efficiency $\alpha$, then the total radiated energy will be:
\begin{equation}
E_{rad} \approx \alpha \frac{M_b}{M_a+M_b} E_{a},
\label{Eq_E_rad}
\end{equation}
where $E_{a}$ is the kinetic energy of mass $M_a$.  

For the CSM interaction scenario, where $M_a$ is the supernova ejecta, and $M_b$ is the circumstellar shell, this approximation is valid since a supernova ejecta's velocity typically reaches $10^4$ km/s while mass previously ejected by stars have velocities ranging from $\sim 10^1$ to $10^3$ km/s.  Also, $E_a\sim 10^{51}$ ergs is approximately the total energy of the supernova, as adiabatic expansion quickly converts the initial deposited energy of the supernova into kinetic form.  The radiative conversion efficiency is typically high, $\alpha \gtrsim 0.5$, at least for optical radiation from thermalized shock material \citep{Moriya2013}.  Thus from equation (\ref{Eq_E_rad}), for a given total system mass and explosion energy, the energy radiated away is linearly proportional to the CSM mass $M_b$.  So although most supernova only radiate 1\% of their total kinetic energy, a large circumstellar mass $M_b$ can substantially recover the supernova energy lost by adiabatic expansion.  Notably, in this toy model, the total radiated energy does \emph{not} depend on the location of the circumstellar mass $M_b$.

Several mechanisms may eject a large mass from the star prior to its death as a supernova.  For example, luminous blue variables (LBVs) are evolved, unstable massive stars, and giant eruptions from LBVs result in dramatically increased mass loss and luminosity, some of which are so extreme that they are initially mistaken for supernova.  These \emph{supernova impostors} are powerful but non-terminal eruptions (i.e. not core collapse), however, there is direct evidence linking at least some LBVs and supernova impostors to actual supernova, e.g. SN 2006jc \citep{Foley2007}, in which the progenitor star is observed to violently erupt only 2 years before its terminal explosion; other examples include SN 2005gl \citep{Gal-Yam2007,Gal-Yam2009a} and possibly SN 2009ip \citep{Mauerhan2012}.  Alternatively, some of the most massive stars with helium core masses between $\sim 40$ to 60 $M_{\odot}$ encounter core instability from the softening of the equation-of-state due to production of electron-positron pairs, which results in explosive burning that is insufficient to fully unbind the star, but can result in a sequence of supernova-like eruptions of shells of matter shortly before the star dies.  The collision of subsequent shells of ejecta can also produce a superluminous supernova, i.e. the pulsational pair-instability SNe \citep{Heger2002,Woosley2007,Chatzopoulos2012}.  Also, the tunneling of wave energy from the core (driven by fusion-luminosity induced convection) into the stellar envelope can lead to extremely large stellar mass loss rates a few years prior to core-collapse \citep{Quataert2012}.  Alternatively, the collective action of winds at different evolutionary stages of the progenitor star can form wind-blown cavities, bordered by a thin, dense, cold shell constituting material swept-up by the winds; the emission of SNe in these wind-blown bubbles have been examined \citep{Chevalier1989,Dwarkadas2005}.

Now, for CSM-interaction powered supernova, the generation of optical emission requires that high densities are still maintained when the SNe ejecta collides with the circumstellar material, usually implying the CSM is relatively near to the star ($\lesssim 10^{15}$ cm).  However, the physical mechanism behind LBV outbursts is not yet known, so there is little theoretical constraint on the timing between the outburst and the supernova afterward; observational constraints so far set the lower limit to 40 days \citep{Ofek2013}, but the delay can be years to decades or longer \citep{Davidson2012}.  As for the pulsational pair-instability mechanism, the interval between pulses can be anywhere from $\sim 1$ week to $>1000$ years \citep{Woosley2007}.  As longer delay times between eruptions imply that subsequent ejecta take longer to catch up to previous ejecta, it is quite possible that the collision between ejecta can occur at larger radii.  As for the CSM shells bordering wind-blown bubbles, they are naturally placed by the duration of winds during late stellar evolutionary stages (e.g. Wolf-Rayet) at least $10^{19}$ to $10^{20}$ cm away from the star.

So, if instead the SNe ejecta encounters a massive CSM shell at larger radii $> 10^{15}$ cm, the shell material is spread thinner, and depending on the CSM shell mass, the resulting shock can be \emph{optically thin}, albeit still dense enough to drive strong emission.  Such an event could still radiate extreme amounts of energy, perhaps comparable to the currently observed superluminous SNe, but the actual optical emission could be quite modest, with the bulk of the radiation instead emitted in X-rays.  

Moreover, in this scenario the bulk of the X-ray emission may come from the forward shock, i.e. from the shocked CSM shell.  This has an important advantage compared with most cases of X-ray emission from young SNe (without a CSM shell), in which the reverse shock is usually denser, and the observed emission is usually attribute to line-cooling emission from the reverse shock running in the SNe ejecta, especially at later times \citep{Chevalier2003}.  An important detriment of the cooling is that the intervening cooled, dense post-shock gas may photoelectrically absorb most of the emission from the reverse shock.   However, even if the forward shock is radiative, and a cool, dense shell forms, this post-forward-shock cool gas will be \emph{behind} the newly shocked CSM with respect to an observer on Earth -- in contrast to the opposite arrangement for the reverse shock.  Thus, for forward shock emission from SN \& CSM shell interactions, only absorption and scattering by the pre-shock CSM is important, and even these go away once the forward shock runs through the CSM shell.   

\citet{Chugai1993} proposed an analogous scenario for the X-ray emission from SN 1986J, in which the emission originates from the forward shock front moving into dense wind clumps, and \citet{Chugai2006} modeled the luminous X-ray emission $\sim 10^{41}$ erg s$^{-1}$ of SN 2001em as interaction of normal SNe ejecta with a dense, massive CSM shell, albeit attributing the observed luminosity to a non-radiative reverse shock.  The evolution of SNe ejecta expanding into a power-law density CSM have been well studied \citep{Chevalier1982,Chevalier1982a}, and simple formulas for its dynamics and emission exist in terms of self-similar solutions; however, these are not applicable for a CSM shell.

In this paper, we consider the forward shock emission from SN ejecta colliding into a CSM shell, and derive simple, general formulas for: (i) the regimes in which the shock will be radiative versus non-radiative, and whether the X-ray luminosity will be powered by free-free emission or line-cooling, and (ii) the approximate luminosity and total energy emitted as a function of the CSM shell mass, distance from the progenitor, and thickness, as well as the SN explosion energy.  We give examples of possible extremely luminous or energetic emission events.

\section{CSM SHELL CHARACTERISTICS}

For the range of masses expelled in LBV eruptions, there have only been two outbursts where we can directly measure the ejected mass -- around 10 $M_{\odot}$ for $\eta$ Car, but only 0.1 $M_{\odot}$ for P Cygni \citep{Smith2011}.  As for pulsational pair-instability events, most pulses eject $\sim$ 1 $M_{\odot}$ shells, but the full range also spans from $\sim$ 0.1 to 10 $M_{\odot}$.  Note that we make a distinction here between eruptive mass loss and wind-driven mass loss, which also occur for LBV-like progenitors of Type IIn SNe.  Model-inferred wind-driven mass loss rates of Type II SNe progenitors are found to range from a few $10^{-2}$ to $10^{-1} M_{\odot}$ yr$^{-1}$ \citep{Kiewe2012}, but smooth winds will result in a $r^{-2}$ density distribution instead of a shell, unless the wind experiences dramatic changes in its mass loss rate or velocity right before stellar demise.  Here we consider the range of masses $M_{CS}$ of the CSM shell in between $10^{-2} M_{\odot} < M_{CS} < 10 M_{\odot}$, and define the dimensionless CSM shell mass $M_{1} \equiv (M_{CS}/1 M_{\odot})$.

For the range of locations for the CSM shell, we consider scenarios where the previously ejected shell of material is at a radius $R_s$ of at least $10^{15}$ to $10^{17}$ cm, which means that even at supernova ejecta velocities of $10^4$ km s$^{-1}$, the interaction event woould not happen until at least several months to several years after the progenitor star's explosion.  Here we consider the radius $R_{CS}$ of the CSM shell in the range of $10^{15}$ cm $< R_{CS} < 10^{19}$ cm, and define the dimensionless CSM shell radius $R_{17} \equiv (R_{CS}/10^{17}{\rm cm})$.

The thickness of the CSM shell is affected by the duration of the mass loss episode.  For many models of episodic mass loss from massive stars, these eruptions occur for 1-10 years every $10^{3-4}$ years, and lose a total of $0.1-10 M_{\odot}$ per episode.   Note that if the mass loss is smooth during the episode, as in the Super-Eddington stead-state continuum driven wind through a porous medium \citep{Shaviv2000,Owocki2004}, then if the heightened mass loss lasts 1 to 10 years with speed 100 km s$^{-1}$, the shell thickness is $3\times 10^{14}$ to $3\times 10^{15}$ cm.  Alternatively, for explosive expulsions of mass, e.g. via the pulsational pair-instability, due to the spread in velocities of the expelled material, the thickness of the CSM shell may be substantial compared to the radius, $\Delta R_{CS} / R_{CS} \sim 1$.   Conversely, for the dense shells bordering wind-blown bubbles, the shells are typically thin $\Delta R_{CS} / R_{CS} \sim 10^{-2}$.  Here we consider the range of thicknesses $\Delta R_{CS}$ of the CSM shell in between $10^{13}$ cm $< \Delta R_{CS} < 10^{17}$ cm, and define the dimensionless CSM shell thickness $\Delta R_{15} \equiv (\Delta R_{CS}/10^{15}{\rm cm})$.

Assuming the CSM shell is spherically symmetric with uniform density, the surface density of the CSM shell is given by $\Sigma = M_{CS} / 4\pi R_{CS}^2$:
\begin{equation}
\Sigma_{CS}= 1.6 \times 10^{-2} \: M_1 \: R_{17}^{-2} \quad {\rm g\:cm^{-2}}.
\end{equation}
The density of the CSM shell will depend on the thickness of the shell, $\rho_{CS} = \Sigma_{CS}/\Delta R_{CS}$, and so we define the electron number density of the CSM shell as
\begin{equation}
n_7 = \frac{n_{CS}}{10^7 {\rm cm^{-3}}} = 0.95 \: M_1 \: R_{17}^{-2} \: \Delta R_{15}^{-1}.
\end{equation}
Note that $n_{CS} \approx 10^{7}$ cm$^{-3}$ corresponds to a mass density of $\rho_{CS} \approx 1.7\times 10^{-17}$ g cm$^{-3}$.  In reality, the CSM shell may be clumpy, but the clumps could be completely crushed and then mixed within the forward shock, making okay the smooth shell approximation at least for the calculation of post-shock dynamics and its X-ray emission \citep{Chugai2006}.  

We only consider regimes where the CSM shell is optically thin, i.e. the optical depth of the CSM shell for electron scattering $\tau = \kappa_{es} \Sigma_{CS} $ is less than unity:
\begin{equation}
\tau = 5.4 \times 10^{-3} \: M_1 \: R_{17}^{-2} < 1.
\label{Equationtau}
\end{equation}
Note that this line-of-sight optical depth does not change even if the post-shock material is compressed and the density rises.  Here we adopt the electron scattering opacity, $\kappa_{es} \approx 0.34$ cm$^2$ g$^{-1}$ at solar abundances.  Once the supernova ejecta collides with the CSM shell, the shock will heat up the shell material, and the temperature right behind the forward shock could reach $10^7 - 10^9$ K, generating $1-100$ keV photons.  But unlike other superluminous Type IIn supernova, in the scenarios considered in this paper, as the shocked material cools and emits free-free radiation, such radiation will generally {\bf not} be re-processed and thermalized by the circumstellar material (to $T\sim 5,000 - 20,000$ K blackbodies temperatures, resulting in optical emission), but instead immediately leak away as X-rays.

\section{THEORY: SIMPLE FORMULAS}
\label{SectionSimpleFormulas}

\subsection{Shock Velocity, Temperature, and Cooling Mechanism}
We assume the pre-shock CSM shell is effectively stationary, i.e. the shock velocity $v_s$ is much greater than the original velocity of the CSM shell.  To find the shock velocity $v_s$ of the forward shock traveling through the CSM shell, we can write the force equation for the shocked CSM shell:
\begin{equation}
\frac{d}{dt}\Sigma_s v_s = P_s(t)
\label{EquationP_s}
\end{equation}
where $P_s(t)$ is the pressure interior to the CSM shell \emph{after} the SN shock hits the shell, and
\begin{equation}
\Sigma_s = \int^{x_s}_0 \rho_{CS} \: dx
\label{EquationSigma_s}
\end{equation}
is the surface density of matter in the shocked CSM shell, and $x_s$ is the distance that the shock has propagated into the CSM shell.  If we make the approximation that the shock velocity is constant, at least within the CSM shell, then we can derive from equations (\ref{EquationP_s}) and (\ref{EquationSigma_s}) that $\rho_{CS} \: v_s^2 = P_s$; that is, the ram pressure pushing back on the shocked CSM shell moving at velocity $v_s$ (thin shell approximation) into the external, stationary CSM equals the post-shock pressure interior to the shocked CSM shell.  Therefore, 
\begin{equation}
v_s = \left( \frac{P_s}{\rho_{CS}} \right)^{1/2}.
\end{equation}
Now, we can approximate the pressure exerted by the SN ejecta immediately before the shock hits the CSM shell as $P_{SN}=(\gamma-1)E_{SN}/V_{SN}=E_{SN}/2\pi R_{CS}^3$; here $V_{SN}$ is the volume interior to the shell, and we assume a $\gamma=5/3$ gas in this paper.  For convenience, we define the dimensionless SN explosion energy $E_{51}\equiv (E_{SN}/10^{51}{\rm erg})$.

However, once the shock hits the CSM shell, the kinetic energy of the flow is converted into thermal energy, and the pressure rises above $P_{SN}$.  By solving the one-dimensional non-radiative gas dynamics of a plane-parallel shock impinging on a density discontinuity, it can be shown that the immediate post-transmitted shock pressure is a factor $\beta$ greater than the pre-transmitted shock pressure, where $\beta$ is a function of the density ratio $\rho_{CS}/\rho_0$ across the density discontinuity at the CSM shell, and $\rho_0$ is the density of material interior to the CSM shell \citep{Sgro1975}:
\begin{eqnarray}
\nonumber
\frac{\rho_{CS}}{\rho_0} 	&=& \frac{3A_r(4A_r-1)}{\{ (3A_r(4-A_r))^{1/2}-5^{1/2}(A_r-1)\}^2}, \\
\beta 							&=& \frac{4A_r-1}{4-A_r}.
\label{Equationbeta}
\end{eqnarray}  
Instead of expressing subsequent equations as a complicated function of $\rho_0$, we use the shock pressure increase factor $\beta$ to parametrize the severity of increase in density at the CSM shell; $\beta$ monotonically increases from 1 to 6, as $\rho_{CS}/\rho_0$ increases from 1 (no obstacle) to $\infty$ (solid wall), with $\beta=$ 2.6, 4.4, 5.4, and 5.8 for $\rho_{CS}/\rho_0 = 10$, $10^2$, $10^3$, and $10^4$.  Also note the immediate post-shock density $n_s$ increases by a factor of $(\gamma+1)/(\gamma-1)=4$ over the pre-shock density $n_{CS}$.  Hence,
\begin{equation}
v_s = \left( \frac{\beta \: P_{SN}}{\rho_{CS}} \right)^{1/2}.
\end{equation}
Note that this is approximately equal to another formula in literature, i.e. $v_s \approx v_{SN}\sqrt{\rho_{SN}/\rho_{CS}}$ \citep{Chugai1993}, where $v_{SN}$ is the SN ejecta velocity.  In this paper, we only consider the forward shock propagating in the CSM shell; but note that after the forward shock overruns the dense CSM shell, the shock will accelerate as it encounters sparser material, and can be modeled using the formalism of \citet{Dwarkadas2005}.

Thus, we can derive the dimensionless shock velocity $v_8\equiv (v_s/10^8{\rm cm\:s^{-1}})$ as:
\begin{equation}
v_8 = 1.00 \: \beta^{0.5} \: E_{51}^{0.5} \: M_1^{-0.5} \: R_{17}^{-0.5} \: \Delta R_{15}^{0.5}.
\end{equation} 

For a strong shock with an infinite Mach number, the conservation of mass, energy, and momentum dictate that the temperature right behind the shock can be related to the shock velocity $v_s$ via $k T = 2[(\gamma-1)/(\gamma+1)]\: m_i \:v_s^2$, where $k$ is Boltzmann's constant, $\gamma$ is the adiabatic index, and $T_i$, $m_i$ are the temperatures and ion masses of each plasma species.  Note that if an electron-proton plasma is maximally out of thermal equilibrium, then $T_e/T_p \sim m_e/m_p \sim 1/1836$; clearly, whether electron-ion energy equipartition has been reached has great consequence to the electron temperature and thus the observational signature.   If the plasma is in full thermal equilibrium, we can use a single temperature $T$ to describe it, with
\begin{equation}
T \approx 1.36 \times 10^7 \: v_8^2 \quad {\rm K}
\end{equation}
Here we have assumed a mean atomic weight $\mu\approx 0.6$ for a fully ionized plasma of solar abundance.

The timescale for electrons and ions to reach equipartition is $t_{eq}\approx 8.4 \: T^{3/2} \: n^{-1} $ in cgs units \citep{Spitzer1962}, implying
\begin{equation}
t_{eq} \lesssim  10^4 \: v_s^3 \: n_7^{-1} \quad {\rm s} 
\end{equation}
where the inequality originates from the fact the post-shock density $n_s\geq 4n_{CS}$ depending on whether the shocked gas further cools and compresses.  As we shall see, for most high luminosity cases, energy equipartition will be reached in a lot less than a day, with $t_{eq}$ being far less than the cooling time $t_{cool}$ or the shock traversal time through the CSM shell $t_{flow}$, and it is mostly safe to assume the electron temperature is the same as the temperature of the ions.

The subsequent luminosity of the shocked hot gas is driven by their mechanism of radiative cooling, captured by the cooling function $\Lambda$.  Even at solar metallicity, the cooling function is a complicated function of temperature.  For simplicity, we approximate its behavior into two regimes \citep{Chevalier1994}:  When $T>4\times 10^7$ K, free-free emission dominates, and $\Lambda \approx 2.5\times 10^{-27} T^{0.5}$ erg cm$^3$ s$^{-1}$, whereas when $10^5$ K $< T \lesssim 4\times 10^7$ K, line emission increases, and $\Lambda \approx 6.2\times 10^{-19} T^{-0.6}$ erg cm$^3$ s$^{-1}$; these are rough fits to the cooling curves calculated by \citet{Raymond1976}.  Hence, we can define a dimensionless cooling function $\Lambda_{-23}=\Lambda(T)/10^{-23}$erg cm$^3$ s$^{-1}$:
\begin{equation}
\Lambda_{-23} = \left\{
\begin{array}{l l}
 0.92 \: v_8 			& \quad \textrm{if $v_8>1.7$ (free-free)}\\
 3.25 \: v_8^{-1.2}	& \quad \textrm{if $v_8<1.7$ (line-cooling)}\\
\end{array} \right.
\label{EquationDimensionlessCoolingRate}
\end{equation}
In reality, the cooling function $\Lambda$ is a function of the emitted photon frequency $\mu$ as well, and a detailed $\Lambda(T,\mu)$ would provide us with an emission spectrum.  We utilize this more involved approach to simulations in \S\ref{SectionSimulation}.

\subsection{Radiative vs Non-radiative Shock}

A radiative shock typically forms when the density of the ambient medium is high enough, such that the emitted radiation affects the dynamics of the gas behind the shock; this occurs when the cooling time $t_{cool}$ is shorter than the hydrodynamical time $t_{flow} \approx \Delta R_{CS} / v_s$:
\begin{equation}
t_{flow} = 0.32 \: \Delta R_{15} \: v_8^{-1} \quad {\rm yr}
\end{equation}  

\begin{figure*}
\centering
\subfloat[$E_{51}=1$, $M_1=1$, $\beta=1$]
{
\includegraphics[width=0.48\textwidth]{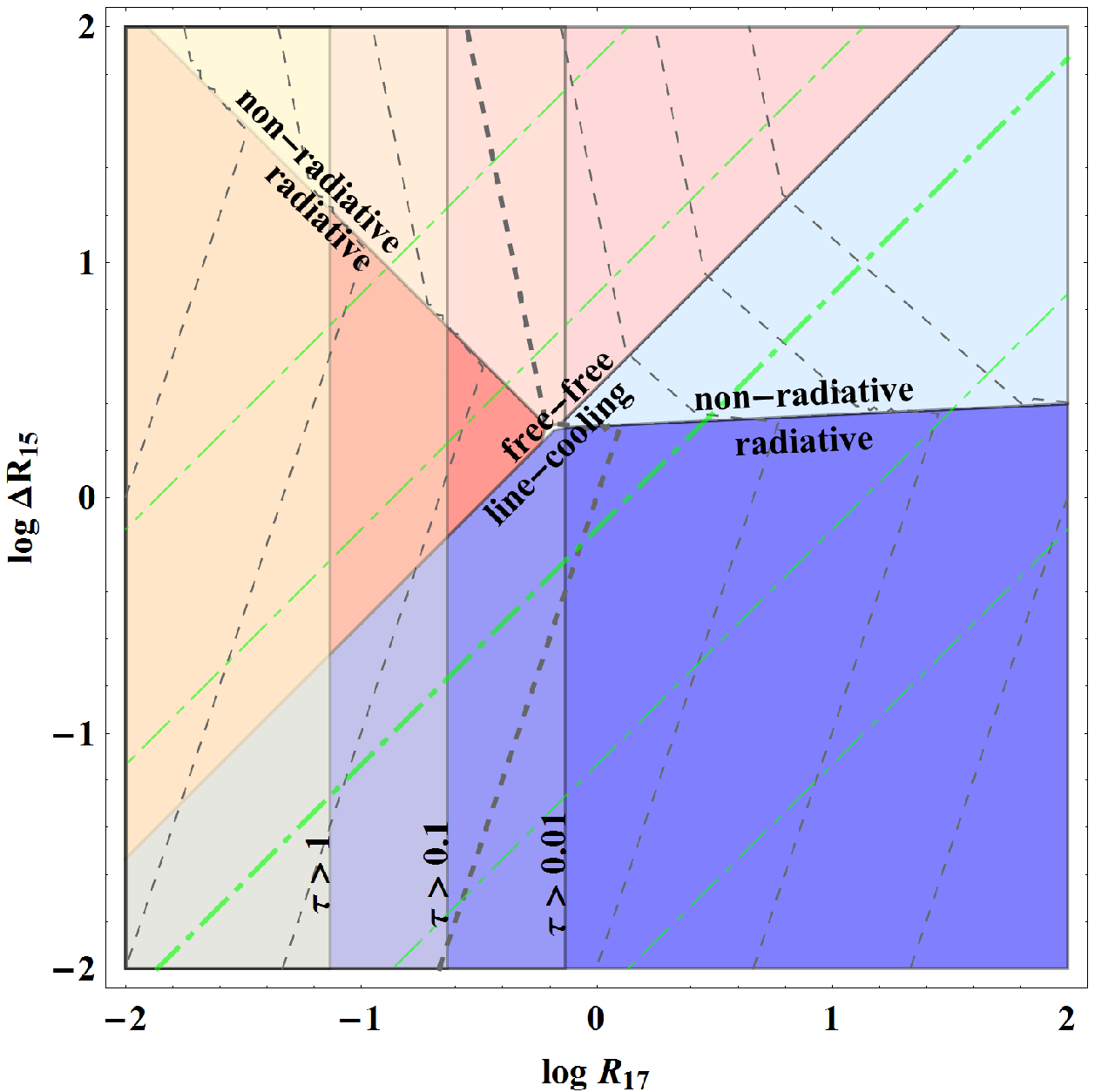}
\label{Figure_lR_ldR_inequalities:E1_M1_beta1}
}
\quad
\subfloat[$E_{51}=1$, $M_1=0.1$, $\beta=1$]
{
\includegraphics[width=0.48\textwidth]{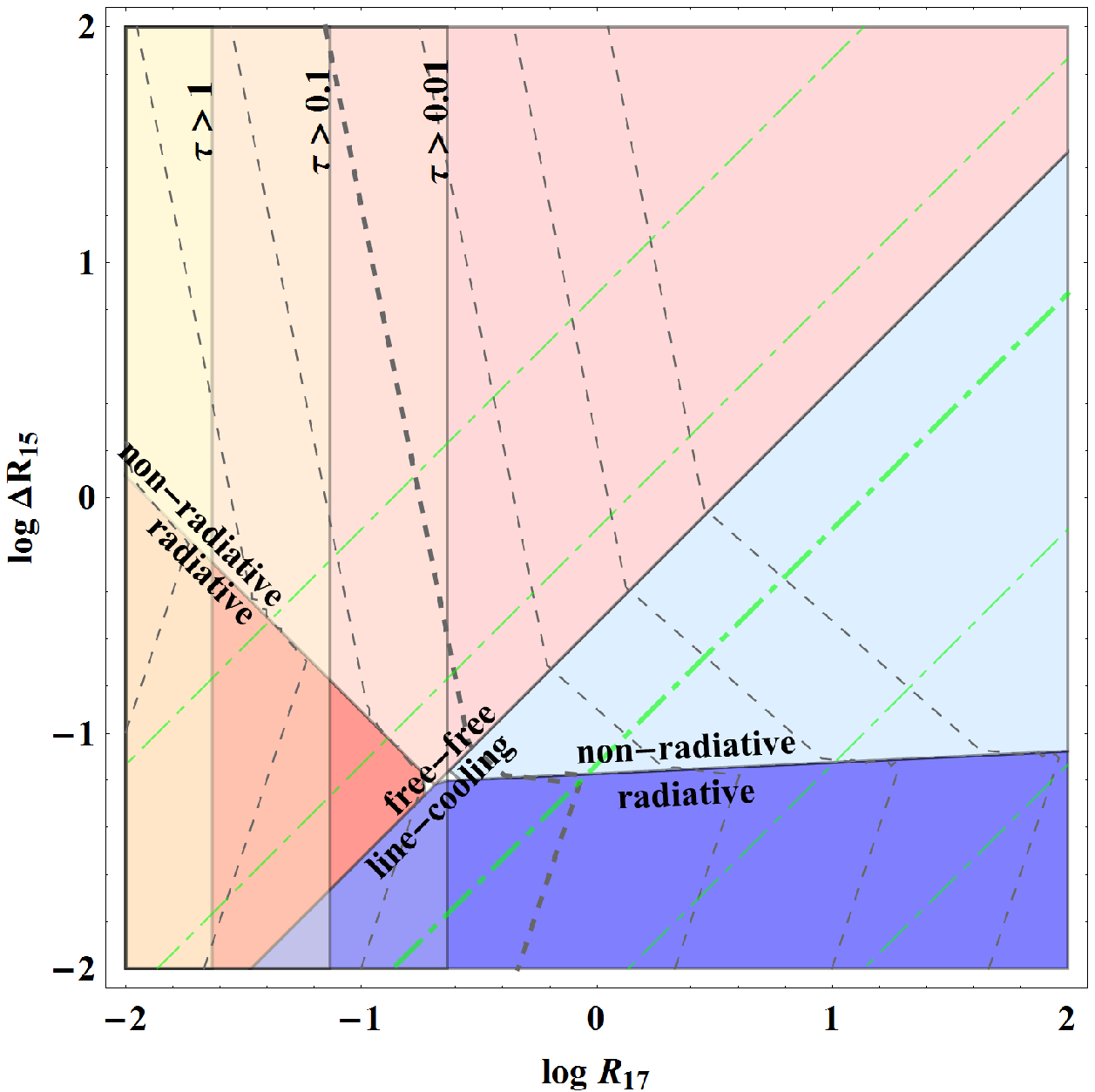}
\label{Figure_lR_ldR_inequalities:E1_M0.1_beta1}
}
\\
\subfloat[$E_{51}=10$, $M_1=1$, $\beta=1$]{
\includegraphics[width=0.48\textwidth]{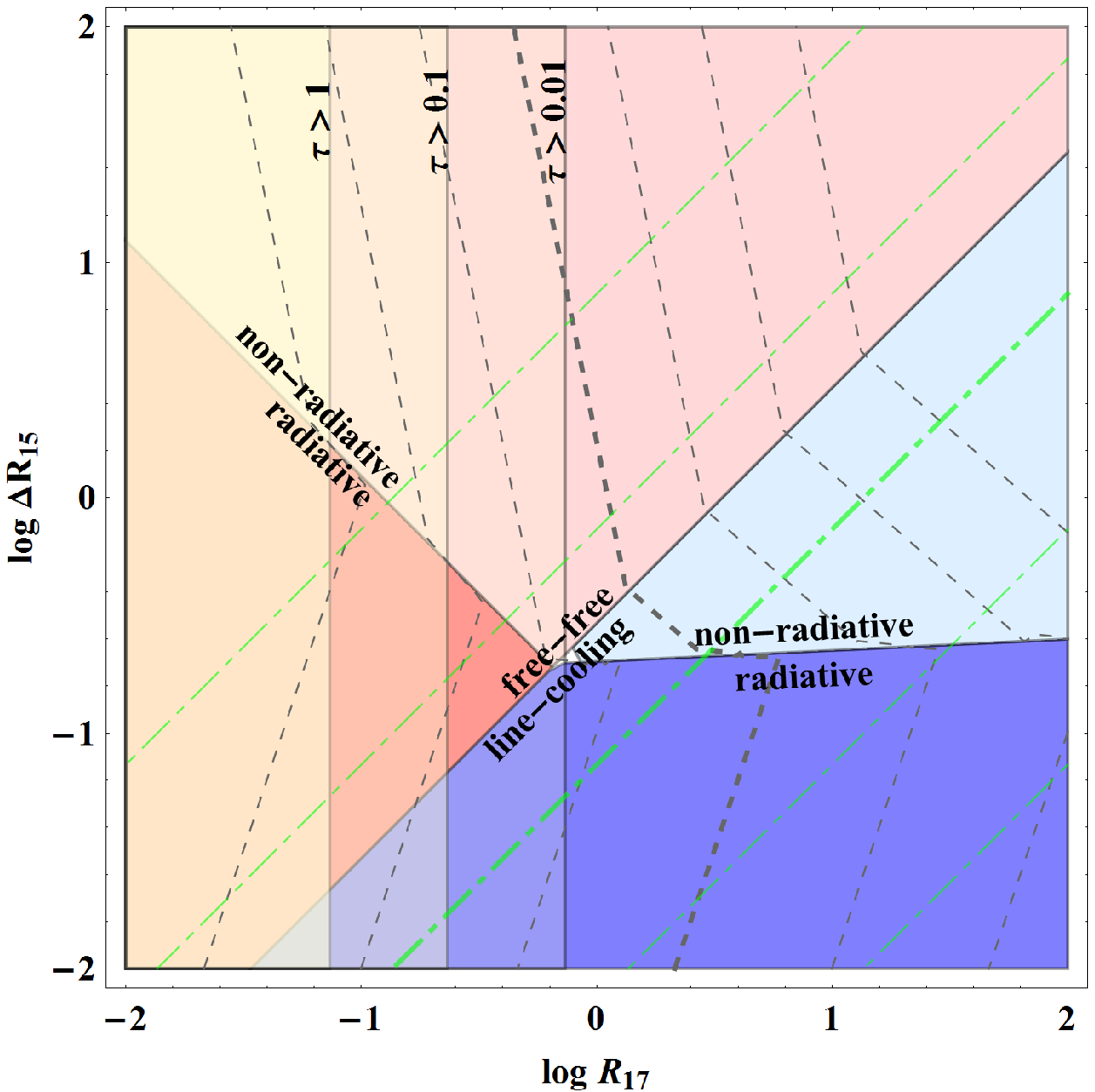}
\label{Figure_lR_ldR_inequalities:E10_M1_beta1}
}
\quad
\subfloat[$E_{51}=1$, $M_1=1$, $\beta=6$]{
\includegraphics[width=0.48\textwidth]{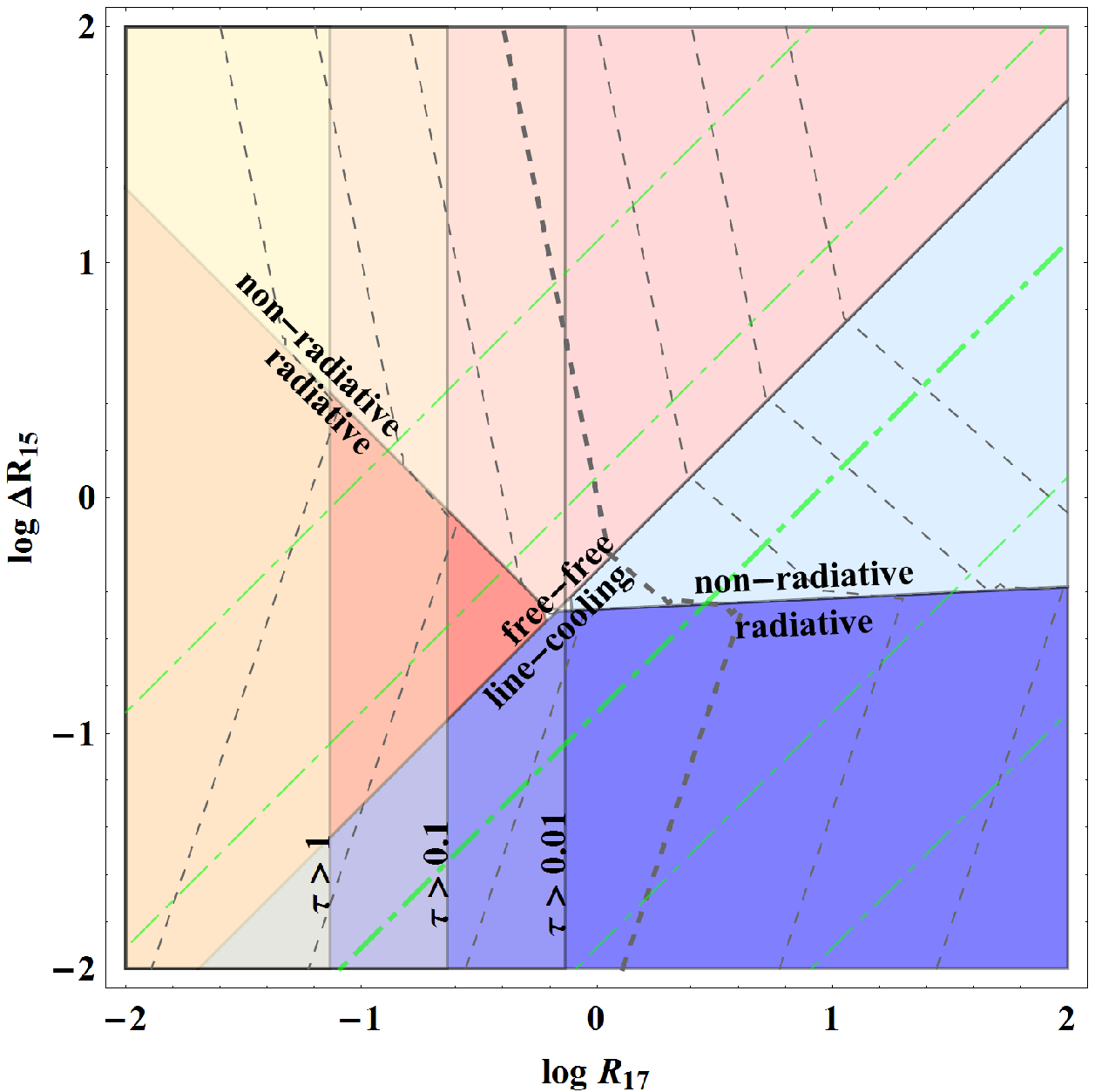}
\label{Figure_lR_ldR_inequalities:E1_M1_beta6}
}
\caption{Emission properties of the shock in the CSM shell, varying the shell radius $R_{17} \equiv (R_{CS}/10^{17}{\rm cm})$ and thickness $\Delta R_{15} \equiv (\Delta R_{CS}/10^{15}{\rm cm})$.  The panels show different choices for the SN explosion energy $E_{51}\equiv (E_{SN}/10^{51}{\rm erg})$, shell mass $M_{1} \equiv (M_{CS}/1 M_{\odot})$, and the shock pressure increase factor $\beta$ (equation (\ref{Equationbeta})).  The red and blue regions cover where the shock is dominated by free-free emission or line-cooling, respectively, in which the darker red and blue regions depict where the shock is radiative.  The overlapping yellow regions show where the electron scattering optical depth along the line of sight is greater than $\tau >$ 0.01, 0.1, and 1, respectively.  The dashed gray lines depict contours of constant X-ray luminosity, with the thicker line indicating where $L_{42} \equiv (L/10^{42}{\rm erg\: s^{-1}}) = 1$; each adjacent line toward the left is more luminous by a factor of 10.  The luminosity roughly increases with $\tau$, but at $\tau>1$ the X-rays start being reprocessed into optical emission instead; hence $10^{44}$ to $10^{45}$ erg s$^{-1}$ is the maximum X-ray luminosity possible.  Similarly, the dot-dashed green lines depict contours of constant shock temperature, with the thicker line indicating where $T = 10^7$ K; each adjacent line in the direction of the red region is hotter by a factor of 10.  Although luminosities up to a few $10^{44}$ erg s$^{-1}$ are possible at $\tau \lesssim 1$, photoelectric absorption is severe (equation (\ref{EquationEtau_pe=1})), and so except for high temperature shocks $T\sim 10^9$ K emitting many $\gtrsim 20$ eV photons, the full luminosity won't be observable until the shock runs through the entire CSM shell.   Similarly, for $10^{43}$ erg s$^{-1}$ pre-absorption luminosities, the early shock emission will be completed obscured unless the temperature reaches $T \gtrsim 10^8$ K.  For the same optical depth (i.e. column density), the highest luminosities are best reached via radiative shocks dominated by free-free emission.
}
\label{Figure_lR_ldR_inequalities}
\end{figure*}

\begin{figure*}
\centering
\subfloat[$E_{51}=1$, $\Delta R_{15}=1$, $\beta=1$]
{
\includegraphics[width=0.48\textwidth]{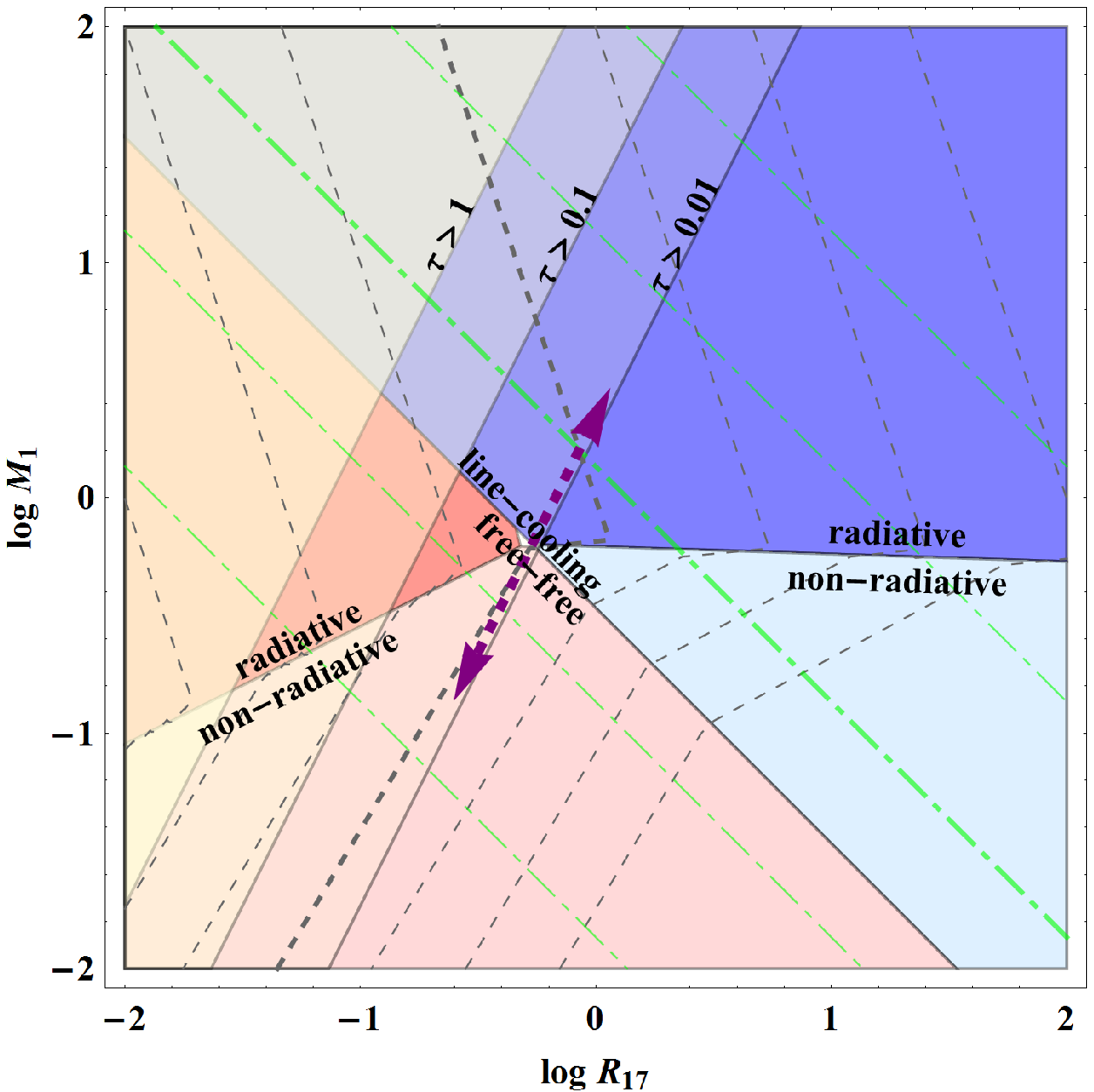}
\label{Figure_lR_lM_inequalities:E1_dR1_beta1}
}
\quad
\subfloat[$E_{51}=1$, $\Delta R_{15}=1$, $\beta=6$]
{
\includegraphics[width=0.48\textwidth]{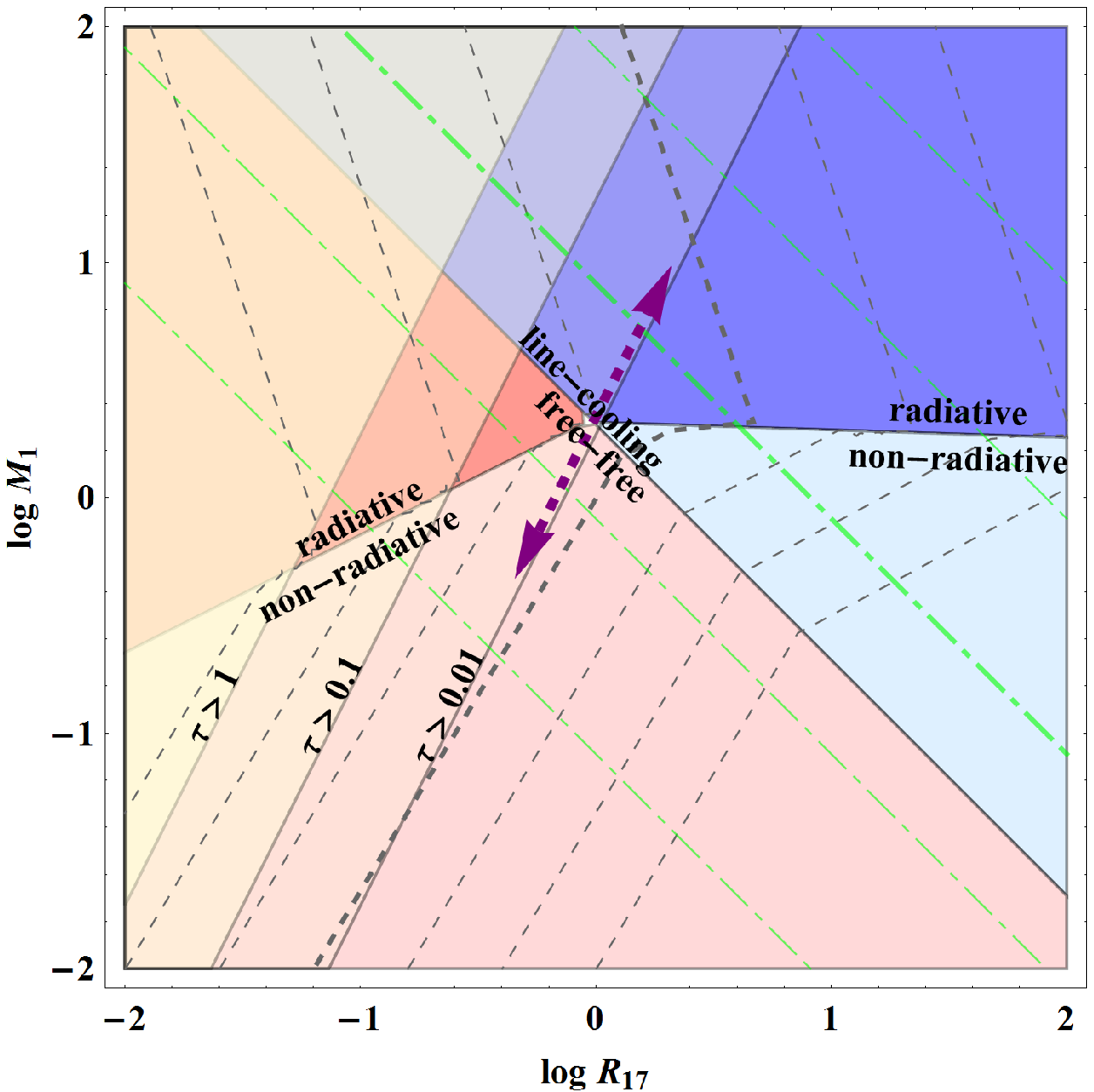}
\label{Figure_lR_lM_inequalities:E1_dR1_beta6}
}
\caption{Emission properties of the shock in the CSM shell, varying the shell radius $R_{17}$ and mass $M_1$.  The color and dashed-line notation is the same as Figure \ref{Figure_lR_ldR_inequalities}.  Note that the intersection point $({R}_{17}^{\prime},{M}_1^{\prime})$ between the free-free vs line-cooling boundary and the radiative vs non-radiative boundary behaves as ${R}_{17}^{\prime}\propto \Delta R_{15}^{1/3} E_1^{1/3}$ and ${M}_{1}^{\prime}\propto \Delta R_{15}^{2/3} E_1^{2/3}$, so that changing $\Delta R_{15}$ or $E_1$ would simply move the intersection point along the white arrow.  Therefore, the intersection point will always lie around where optical depth $\tau=0.01$ (as $\tau$ is independent of $\Delta R_{15}$ and $E_1$), thus the free-free \& radiative regime will always have $\tau>0.01$, with a pre-shock CSM shell column density $N_H > 1.8\times 10^{23}$ cm$^{-2}$.}
\label{Figure_lR_lM_inequalities}
\end{figure*}

The cooling time of a gas element in a shock can be calculated as the ratio between the thermal energy density $\epsilon = 3/2n_s kT$ and the cooling rate per unit volume $\hat{\Lambda}=n_s^2 \Lambda$ \citep{Franco1993}:
\begin{equation}
t_{cool}=\frac{\epsilon}{\hat{\Lambda}}\approx \frac{3kT}{2n_s\Lambda(T)},
\label{Equationt_cool}
\end{equation}
where $n_s$ is the immediate post-shock density.  Thus, depending on the shock temperature,
\begin{equation}
t_{cool} = \left\{ \begin{array}{l l l}
 0.24 \: v_8 \: n_7^{-1} 			& {\rm yr}		& \quad \textrm{if $v_8>1.7$ (free-free)}\\
 0.07 \: v_8^{3.2} \: n_7^{-1} 	& {\rm yr} 	& \quad \textrm{if $v_8<1.7$ (line-cooling)}\\
\end{array} \right.
\label{Equationtcool}
\end{equation}
Thus, the condition for a radiative shock $t_{cool} < t_{flow}$ can be expressed as:
\begin{equation}
\left\{
\begin{array}{l l l}
 v_8^2 		&< 1.31 \: n_7 \: \Delta R_{15}		& \quad \textrm{if $v_8>1.7$ (free-free)}\\
 v_8^{4.2} 	&< 4.61 \: n_7 \: \Delta R_{15} 		& \quad \textrm{if $v_8<1.7$ (line-cooling)}\\
\end{array} \right.
\label{Equationradiativeshockcondition}
\end{equation}
We plot the dependence of these different regimes on the CSM shell mass $M_1$, radius $R_{17}$, and thickness $\Delta R_{15}$ in Figures 
\ref{Figure_lR_ldR_inequalities}
and \ref{Figure_lR_lM_inequalities},
noting that the transition between regimes is much smoother than depicted.

\subsection{Luminosity and Total Energy Emitted}

\subsubsection{Non-radiative shock}

The X-ray luminosity of a non-radiative shock heated plasma can be calculated as $L = {\rm EM} \times \Lambda$, where EM is the emission measure, and $\Lambda$ is the cooling function.  The emission measure for the fully shocked CSM shell can be calculated as the emission volume $V_{CS}/4$, which is the CSM shell volume $V_{CS} =4\pi R_{CS}^2 \Delta R_{CS}$ compressed by the shock, multiplied by the square of the post-shock density $n_s = 4n_{CS}$, assuming that the density is uniform throughout.  Thus,
\begin{equation}
{\rm EM} = 4.50\times 10^{64} \: M_1^2 \: R_{17}^{-2} \: \Delta R_{15}^{-1}  \quad{\rm cm^{-3}}.
\end{equation}
Combined with the cooling rate at different shock velocities/temperatures in equation (\ref{EquationDimensionlessCoolingRate}), we can find the non-radiative luminosity as a function of system parameters, expressed in terms of a dimensionless X-ray luminosity $L_{42} \equiv (L/10^{42}{\rm erg\: s^{-1}})$ as follows.

When $v_8>1.7$, the luminosity of the non-radiative shock set by free-free emission (thermal bremsstrahlung) is
\begin{equation}
L_{42} = 0.42 \: \beta^{0.5} \: E_{51}^{0.5} \: M_1^{1.5} \: R_{17}^{-2.5} \: \Delta R_{15}^{-1.5}.
\label{EquationL42nonradiative_free}
\end{equation}

When $v_8<1.7$, the luminosity of the non-radiative shock set by line-cooling is
\begin{equation}
L_{42} = 1.46 \: \beta^{-0.6} \: E_{51}^{-0.6} \: M_1^{2.6} \: R_{17}^{-1.4} \: \Delta R_{15}^{-1.6}.
\label{EquationL42nonradiative_line}
\end{equation}

Assuming no other energy loss mechanism, we can naively estimate the total energy emitted as $L_X t_{cool}$; however, since non-radiative shocks can have extremely long cooling times, expansion of the shocked CSM shell can convert its thermal energy back into bulk kinetic form, instead of eventually emitting the energy as radiation.  The shocked CSM shell expansion time scale is roughly:
\begin{equation}
t_{exp} = 31.7 \: R_{17} \: v_8^{-1} \quad{\rm yr}.
\end{equation}
This is the time it takes the shocked shell to double in radius, and lose half its energy via $PdV$ work.  Therefore, we estimate the total energy released via:
\begin{equation}
E_X = L \times \min( t_{cool}, t_{exp} ).
\end{equation}

\subsubsection{Radiative shock}

An important difference between a radiative shock and a non-radiative shock is that the former can increase the density drastically by a factor of  $f_n\gg 4$.  Immediately downstream from the shock, the Rankine-Hugoniot jump conditions are still valid, and the density has been compressed by only a factor of 4.  However, as the shocked gas radiates energy away further downstream, its temperature drops precipitously, and its density increases to compensate and keep the total pressure constant.  At approximately a cooling length $L_{cool}= v_s t_{cool}$ away, the shocked gas condenses into a cold, dense shell; the density increase is usually limited to a factor of $\sim 100$ by magnetic pressure.   

Therefore, in calculating the luminosity of a radiative shock, the emission measure will never reflect the entire shocked CSM shell volume, as material one cooling length $L_{cool}$ downstream from the shock will have cooled `completely' and no longer contribute X-ray emission.  The emission measure can thus be approximated as the emission volume $4\pi R_{CS}^2 L_{cool}/f_n$ (accounting for compression) multiplied by the post-shock density squared $n_s^2 = f_n^2 n_{CS}^2$.  Using equation (\ref{Equationt_cool}), and noting that the average kinetic energy $3/2kT\approx 1/2m_p v_s^2$ per particle, we find that the kinetic energy of the explosion is converted to radiation at a rate:
\begin{eqnarray}
\nonumber
L 	&=& 2\pi \: R_{CS}^2 \: \rho_{CS} \: v_s^3 \\
  	&=& 0.99\times 10^{42} M_1 \: \Delta R_{15}^{-1} \: v_8^3 \quad {\rm erg \: s^{-1}},
\end{eqnarray}
where $\rho_{CS}$ is the pre-shock density.  Hence, the luminosity of a radiative shock is
\begin{equation}
L_{42} = 0.99 \: \beta^{1.5} \: E_{51}^{1.5} \: M_1^{-0.5} \: R_{17}^{-1.5} \: \Delta R_{15}^{1.5}.
\label{EquationL42radiative}
\end{equation}

Note that because of occultation by the interior SN ejecta, only half of the above X-ray luminosity typically escapes to the observer.  However, since the X-ray emission from the radiative forward shock will emit in all directions, i.e. both toward the observer, and backward into the cooled material behind the forward shock front, the latter cold dense material could reprocess the X-ray, resulting in concurrent optical emission.

The total energy released in X-rays can be approximated as $E_X \approx L \times t_{flow}$.  However, if photoelectric absorption is severe (see next subsection), and none of the emitted X-rays escape until the shock front reaches the end of the CSM shell, the total energy emitted observable in X-rays may only be $E_X \approx L \times t_{cool}$.

\subsection{Scattering and absorption with the pre-shock CSM shell}

We first emphasize that, after the shock runs through and superheats the entire CSM shell, many effects that decrease the transmitted X-ray flux become irrelevant, as there is no intervening material left from the initially cold CSM shell to absorb or scatter X-ray photons.  This is implicitly assumed in our luminosity formula for non-radiative shocks in equations (\ref{EquationL42nonradiative_free}) and (\ref{EquationL42nonradiative_line}), which consider the entire volume of the shocked CSM shell in the emission measure.  However, it is useful to understand photon interactions with the pre-shock CSM, to characterize the observable emission of the forward shock at early times as it just begins to propagate through the CSM shell.

The column density $N_H = \Sigma/m_p$ is given by,
\begin{eqnarray}
\nonumber
N_H 	&=& 9.6 \times 10^{22} \: M_1 \: R_{17}^{-2} \quad {\rm cm^{-2}} \\
	 	&=& 1.8 \times 10^{25} \: \tau \quad {\rm cm^{-2}}.
\end{eqnarray}
We express $N_H$ as a function of the electron scattering optical depth $\tau$ (from equation (\ref{Equationtau})), for ease of comparison in Figures \ref{EquationL42nonradiative_free} and \ref{EquationL42nonradiative_line}.  The effective cross-section for photoelectric absorption is $\sigma(\lambda)\approx 2.2\times 10^{-25} \lambda^{8/3}$ cm$^2$ for a solar composition gas, where $\lambda$ is the X-ray photon wavelength in units of \AA.   This implies the threshold photon energy for photoelectric absorption is
\begin{eqnarray}
\nonumber
E(\tau_{pe}=1)	&\approx& 1.2 \left[ \frac{N_H}{10^{22} {\rm cm^{-2}}} \right]^{3/8} \quad {\rm keV} \\
					&\approx& 20 \: \tau^{3/8} \quad {\rm keV}
\label{EquationEtau_pe=1}
\end{eqnarray}
below which we can assume the observed spectrum is suppressed \citep{Chevalier2003}.  Note that the dense CSM shell is likely to be fragmented and clumpy, due to Rayleigh-Taylor instabilities.  For a fixed shell mass, a non-uniform, clumpy shell will typically result in less overall absorption compared with the uniform density shell we have assumed in this paper; so our inferences regarding photoelectric absorption are somewhat pessimistic.  In any case, for column densities $N_H\geq 10^{24}$ cm$^{-2}$, X-rays $<10$ keV are absorbed, and one needs to observe the source at 10-100 keV.  If the column density increases to $N_H\approx 10^{25}$ cm$^{-2}$, primary X-rays up to several tens of keV are absorbed.  So in order to observe high X-ray luminosities before the shock has passed through the CSM shell, simply requiring the optical depth $\tau \lesssim 1$ of the CS is grossly insufficient, unless the shock temperature is high $T\sim 10^9$ K ($v_8\sim 10$), or that the shock luminosity itself can ionize the CSM shell.  

Assuming that photoionization is determined by the current X-ray luminosity, we can define an ionization parameter $\xi = L/nR^2$ in cgs units \citep{Tarter1969}:
\begin{equation}
\xi = 10 \: L_{42} \: M_1^{-1} \: \Delta R_{15},
\label{EquationIonizationParameter}
\end{equation}
which determines the ratio of photon flux to particle number density for a fixed temperature of the X-ray source.  Typically, for shock temperatures around $T\sim 10^8$ K, the intermediate elements (such as C, N, O) are fully ionized when $\xi>10^2$, but ionizing the heavier elements such as Fe require $\xi\geq 10^3$.  The medium is completely ionized once $\xi\sim 10^4$ \citep{Chevalier2012}, and there is no photoelectric absorption regardless of high column densities.  These conditions are slightly modified for higher energy photons from $T\sim 10^9$ K shocks, as they are more effective at ionizing atoms with higher atomic numbers.

Also, Compton scattering can affect the escape of high-energy photons, as the inelastic scattering of photons transfers energy from the photon away to the scattered electron, increasing the photon wavelength by $\sim h/m_e c$ and thus decreasing the photon energy by $\Delta E \sim E^2/m_e c^2$.  Since the number of scatterings is $\sim \tau_{es}^2$, above a cutoff energy $E_{max} = \Delta E \tau_{es}^2$ the photon energy will be entirely depleted via Comptonization.  Therefore, the cutoff energy can be approximated via $E_{max} \sim m_e c^2/\tau_{es}^2$.  But since the pre-shock optical depth $\tau_{es} < 1$ for the scenarios considered in this paper, most of our X-ray emission at photon energies $\ll 0.5$ MeV will not suffer Compton degradation.

\section{POSSIBLE LUMINOUS EVENTS}
\label{SectionLuminousEvents}

Next, we discuss possible configurations of the CSM shell that give rise to luminous X-ray emission $\gg 10^{42}$ erg s$^{-1}$, i.e. more luminous than any X-ray transient observed so far attributed to SN ejecta interactions with the CSM.  Conservatively, we use only typical SN explosion energies of $10^{51}$ erg (despite the fact that many optically superluminous SNe have been inferred to have $>10^{52}$ erg explosion energies), and we also assume that the post-transmitted shock pressure does not increase substantially, i.e. $\beta\approx 1$.  In actuality, the density jump from the CSM shell interior to the shell itself can be very large, and $\beta \approx 5$ - 6 is quite possible; therefore, our cautious estimates may have underestimated the shock velocity by a factor of $\beta^{0.5}$, the shock temperature by a factor of $\beta$, and the luminosity of radiative shocks by a factor of $\beta^{1.5} \sim 10$!

Generally, in our parameter space, CSM shells that give rise to the most luminous X-rays $L\gtrsim 10^{44}$ erg s$^{-1}$ have radii $R_{CS} \leq 10^{16}$ cm; this is because higher luminosities are reached at higher shell densities, with the largest luminosities being reached when the Thomson scattering optical depth $\tau$ is very close to 1 but not greater.  Luminosities above $10^{43}$ erg s$^{-1}$ are generally dominated by free-free emission.  Energetic, long duration events can be found either for radiative shocks in moderately thick shells, or non-radiative shocks with very large emission volumes.  We give specific examples below, and briefly discuss their observational signature.

\subsection{Long duration events}

\subsubsection{Super-luminous \& energetic free-free emission}
\label{SectionSuperLuminousAndEnergetic}

In this example, the CSM shell has a mass of $1 M_{\odot}$, radius of $10^{16}$ cm, and thickness of $10^{16}$ cm, reaching a pre-shock density of $10^{8}$ cm$^{-3}$.   Electron-ion energy equipartition is reached in $t_{eq} \sim 10$ days or less, and the unabsorbed luminosity from the radiative shock attains $\sim 10^{44}$ erg s$^{-1}$ for about 100 days, liberating a majority of the SN explosion energy; this is our X-ray analogue of optically super-luminous SN!  In this extreme case, the shock is essentially trapped in the CSM shell; the kinetic energy of the SN ejecta will be radiated away, and this infant supernova remnant, less than one year of age, will go directly to the radiative phase, avoiding the Sedov phase.

The initial column density is a staggering $10^{25}$ cm$^{-2}$, which if neutral can absorb all X-rays below $\sim 16$ keV.  However, not only does the fast $v_s\sim 10^4$ km s$^{-1}$ shock emit photons $\gtrsim 20$ keV, but the large ionization parameter $\xi \sim 10^4$ implies that the early shock luminosity will quickly and completely ionize the remaining unshocked CSM shell material, warding off photoelectric absorption.  Therefore, the $\sim 10^{44}$ erg s$^{-1}$ intrinsic luminosity will be observable for most of this event's 3 month duration.

Non-radiative shocks can generate luminous events too.  For example, for a shell mass of $0.2 M_{\odot}$, radius of $5\times 10^{15}$ cm, and thickness of $2\times 10^{15}$ cm, when the shock escapes the shell, a peak X-ray luminosity of $5\times 10^{43}$ erg s$^{-1}$ is attained, after which the entire shocked CS emits and cools for 1 month.  These adiabatic shocks can have long equipartition times; here $t_{eq}\sim 9$ days is not an issue, but other luminous, non-radiative shocks could have equipartition times significantly exceeding the cooling time.

Regardless of whether the shock is radiative or not, we find that almost all super-luminous (i.e. $>10^{43}$ erg s$^{-1}$) and long-duration (i.e. $\gg 1$ day) events have fast, hot shocks dominated by free-free emission.  Less luminous versions may have already been seen, e.g. SN 2010jl \citep{Chandra2012}.

\subsubsection{Luminous line-emission from radiative shocks}

We consider a massive $5 M_{\odot}$ CSM shell with a radius of $2\times 10^{16}$ cm, and thickness of $2\times 10^{15}$ cm.  The pre-shock density is quite high, $n_{CS} = 6\times 10^8$ cm$^{-3}$, but due to the large radius, the pressure from the supernova is spread over a larger area, so that the shock velocity is only 1,400 km s$^{-1}$, and thus the shock temperature $T\sim 3\times 10^7$ K is much cooler than the previous super-luminous examples, resulting in softer X-ray photons of a few keV.  The resulting radiative shock produces a respectable pre-absorption luminosity of roughly $7\times 10^{42}$ erg s$^{-1}$ for half a year, converting $10\%$ of the SN explosion energy into radiation.  However, the column density is $10^{25}$ cm$^{-2}$ like before, but now the ionization parameter is only $\xi \sim 30$, and can only partially ionize the intermediate elements.  Therefore, during most of the 160 days it takes for the radiative shock to traverse the CSM shell, the X-ray flux will suffer heavy photoelectric absorption, and we will not see a rise in luminosity until the shock nears the end of the shell, after which the shock will cool in a matter of days.

Hence, these intrinsically luminous radiative shocks dominated by line-emission may have long underlying durations, but their actual observable durations are typically short. 

\subsubsection{Modest line-emission from non-radiative shocks}

In this example, the CSM shell has a mass of $0.5 M_{\odot}$, radius of $2\times 10^{17}$ cm, and thickness of $10^{15}$ cm, reaching a pre-shock density of $10^{6}$ cm$^{-3}$.  The optical depth is only $\tau=7\times 10^{-4}$, i.e. the column density is $10^{22}$ cm$^{-2}$; this is much less than the previous examples, but the shock temperature here is only $1.4\times 10^7$ K, so the X-ray emission is soft, and much of it will still be absorbed.  Therefore, the peak luminosity of $9\times 10^{40}$ erg s$^{-1}$ will not be observable until the shock traverses the entire shell.  However, it takes the shocked shell material over half a year to cool, so the shocked CSM shell will emit for this length of time even after the shock has left the shell, making it easily observable.

\subsection{Short duration events}

\subsubsection{Super-luminous flares?}

If the CSM shell has a mass of $0.05 M_{\odot}$, radius of $2\times 10^{15}$ cm, and thickness of $10^{14}$ cm, reaching a pre-shock density of $10^{10}$ cm$^{-3}$,  the X-ray luminosity from the resulting shock reaches a staggering $5\times 10^{44}$ erg s$^{-1}$, but only lasts for 1 day, liberating $\sim 5\%$ of the SN explosion energy.  The shock velocity reaches $10^4$ km s$^{-1}$, and $t_{eq}$ is only 1/10 the duration of this event, so electron temperatures of $10^9$ K will be reached rapidly; this proposed class of events will generally produce extremely hard X-rays with a Bremsstrahlung spectrum.

In reality, the spherical symmetry of the CSM shell is likely to be broken.  For instance, if the radii of the CSM shell at different locations varies by a factor of 2, the emission would be spread over a month, reaching less extreme luminosities of $\sim 10^{43}$ erg s$^{-1}$.

\subsubsection{Luminous cool flares?}

It is possible for a radiative shock to generate a luminous X-ray flare powered by line-emission, albeit at lower luminosities than before.   For example, consider a CSM shell with mass $0.2 M_{\odot}$, radius $2\times 10^{16}$ cm, and thickness $10^{14}$ cm; the pre-shock density is still high $10^{9}$ cm$^{-3}$, but the shock velocity is only 1,600 km s$^{-1}$, resulting in a characteristic photon energy of only $\sim 3$ keV.  The luminosity reached for these events can be $\sim 10^{43}$ erg s$^{-1}$, however, the column density is typically large $\gtrsim 5\times 10^{23}$ cm$^{-2}$, with the ionization parameter $\xi<10^2$ insufficient to ionize the unshocked shell material.  Hence, the full luminosity can be observed for only a few days, when the shock reaches the end of the CSM shell.

\section{SIMULATION}
\label{SectionSimulation}

To investigate the time evolution of the supernova shock interacting with
the ejected circumstellar shell, we performed hydrodynamical simulations including
a time dependent ionization calculation.  We focus on scenarios where the shock in the CSM shell is luminous and adiabatic, and
do not address the regime where
strong radiative cooling is important. We have employed the numerical
hydrodynamics code VH-1 \citep[e.g.][]{blondin93} using the nonequilibrium 
ionization calculation similar to that discussed in \citet{patnaude09} but without
the diffusive shock acceleration calculation. 

We model the supernova ejecta as a powerlaw in velocity ($\rho_{\mathrm{ej}}$
$\propto$ $v^{-n_{ej}}$), and take into account its interaction with a circumstellar 
wind within the CSM shell. Except for one model,
we fix the supernova ejecta mass at 4M$_{\sun}$, energy at 2$\times$10$^{51}$ erg,
and powerlaw index at $n_{ej} = 10$, and model the CSM shells
with a range of masses (0.1 -- 1.0M$_{\sun}$), and thicknesses (10$^{14}$ -- 
10$^{15}$ cm), but fix the distance at 10$^{16}$ cm.  The circumstellar wind is derived from a 
progenitor mass-loss rate of $\dot{M}$ = 2$\times$10$^{-5}$ M$_{\sun}$ yr$^{-1}$
with a wind velocity of 10 km s$^{-1}$.  Shells at distances
much greater than 10$^{16}$ cm would produce X-ray emission at later times
than considered here. We model the interaction between 10 days and 0.8 yr after
the supernova. The upper limit on the timescale allows for the shock
to fully traverse the CSM shell.

\begin{figure*}
\begin{minipage}[c]{1.0\textwidth}
\includegraphics[width=0.5\textwidth]{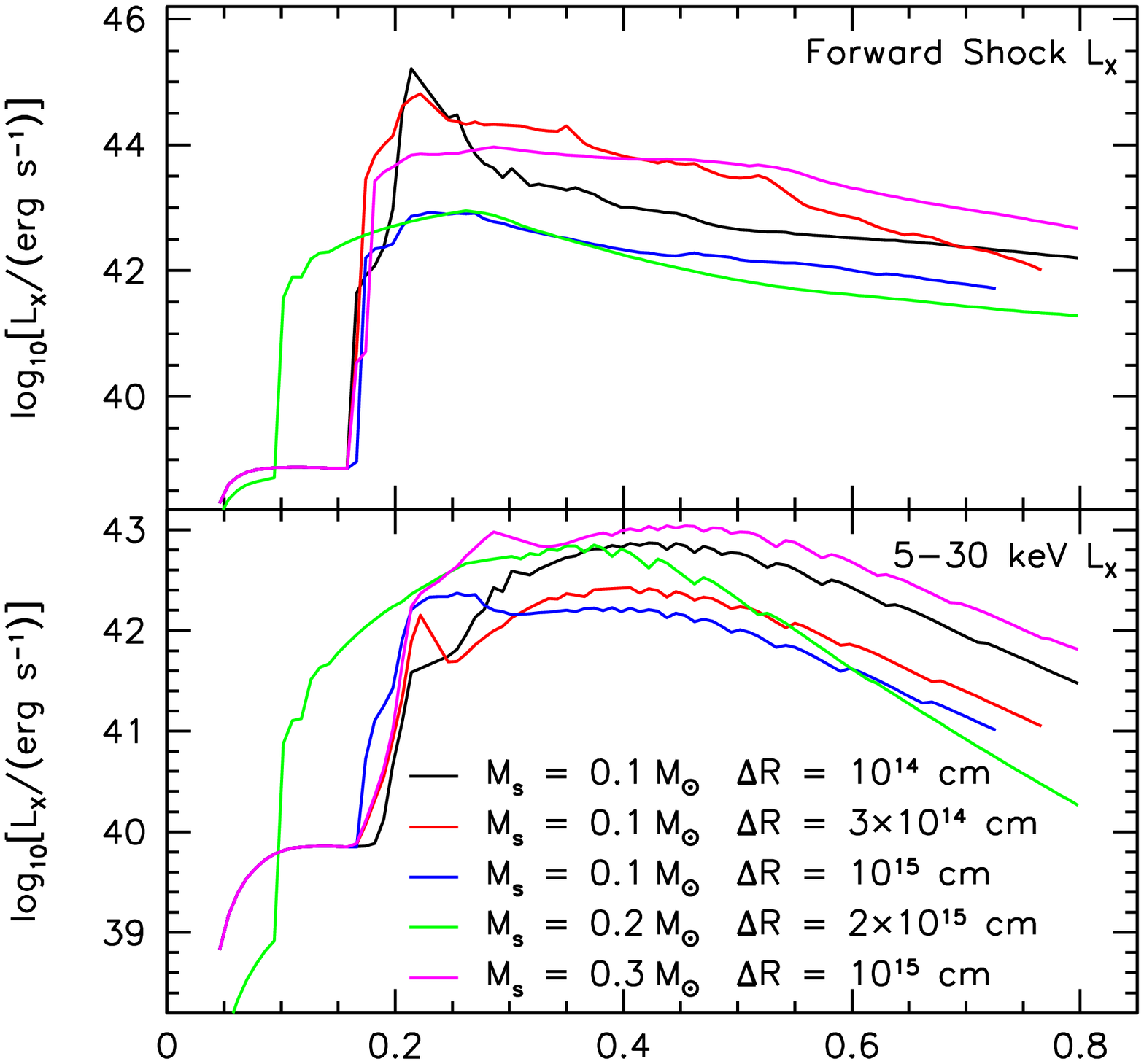}
\includegraphics[width=0.5\textwidth]{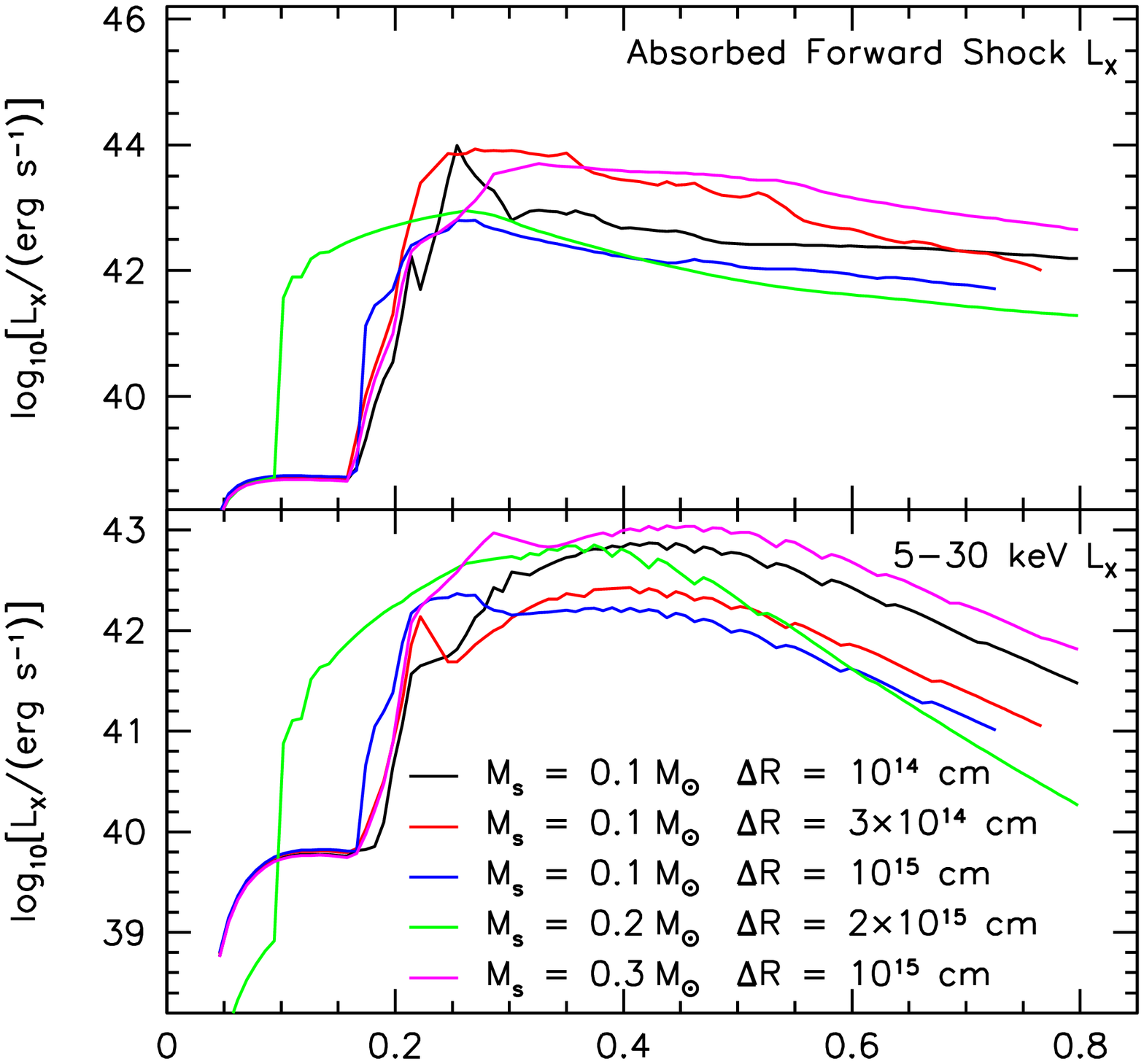}
\end{minipage}
\caption{{\it Left}: In the upper panel, we show the X-ray emission as a function of
time from material swept up by the supernova forward shock. In the lower panel, we plot
the 5--30 keV emission as a function of time. This includes contributions from the forward
shock as well as a negligible contribution from reverse-shocked ejecta. {\it Right}: The
same as in the left panel, except that we include photoelectric absorption from unshocked
circumstellar material, including the unshocked CSM shell.}
\label{fig:model_ltc}
\end{figure*}

We compute the 0.5 -- 30.0 keV thermal X-ray emission as a function of time
to compare against the results depicted in Figures \ref{Figure_lR_ldR_inequalities} and 
\ref{Figure_lR_lM_inequalities}, as well as some of the adiabatic shock scenarios described in 
Section~\ref{SectionLuminousEvents}. We plot the unabsorbed and absorbed 
luminosity versus time for several 
models in Figure~\ref{fig:model_ltc}.  We discard any models where the total radiated 
X-ray luminosity exceeds the supernova kinetic energy; these models have strong radiative 
shocks, outside the regime of validity for our simulation code.  The luminosities seen in 
Figure~\ref{fig:model_ltc} are in general agreement with the predictions from the simple theory 
of Section~\ref{SectionSimpleFormulas}.

As shown in Figure~\ref{fig:model_ltc}, the models chosen here are indeed able to attain 
luminous X-ray emission with L$_{X}$ $\approx$ 10$^{42-44}$ erg s$^{-1}$, once the 
blastwave hits the shell.  Most models show a fast rise in emission once the blastwave impacts the shell, followed by a 
slow decline.

We also plot in Figure~\ref{fig:model_ltc} the super-luminous non-radiative shock 
discussed in the 3rd paragraph of Section \ref{SectionSuperLuminousAndEnergetic}. Here 
we have a low mass ejected CSM shell closer to the star, but also a much wider shell. This results in
a longer rise time in emission (once the blastwave hits the shell, at around $0.15$ yr). 
This  model also contains half the explosion energy as the other models, and the blastwave 
transits across the shell for a longer period of time.

\begin{figure}
\includegraphics[width=0.5\textwidth]{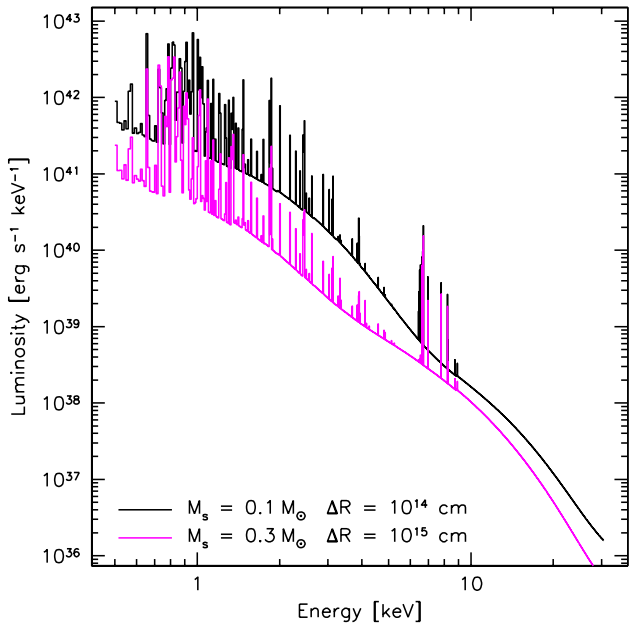}
\caption{0.5--30 keV thermal X-ray emission for two models shown in 
Figure~\ref{fig:model_ltc}, when the shocks break out of the CSM shells at approximately 0.4 and 0.55 yr, respectively. The black curve
corresponds to the model with M$_{s}$ = 0.1M$_{\odot}$ and $\Delta$R$_{s}$ = 10$^{14}$ 
cm, while the magenta curve corresponds to the model with M$_{s}$ = 0.3M$_{\odot}$ and
$\Delta$R$_{s}$ = 10$^{15}$ cm.}
\label{fig:spec}
\end{figure}

Our model also computes the detailed thermal X-ray emission out to 30 keV. In 
Figure~\ref{fig:spec}, we show the X-ray emission at the point when the shocks break
out of the circumstellar shells. The overall normalization, spectral lines, and
line ratios differ significantly between these two models. The shape of the underlying
continua also show differences, particularly above 10 keV where the model with the
thicker shell shows a steeper spectrum at high energies (though appears flatter than
the model with the thin shell at low energies). While the spectral resolution and 
throughput of current X-ray observatories may not be able to discriminate between these
models, high spectral resolution missions such as {\it Astro-H} may be able to.

\section{DISCUSSION}
\label{SectionDiscussion}

Our simple formulas are in rough agreement with other predictions in literature.  Adapting our formulas with a filling factor for clouds in the wind-blown CSM of SN 1986J \citep{Chugai1993}, we arrive at similar luminosities and shock temperatures as observed.  For SN 1987A, our model agrees exactly with the luminosity $L=4\times 10^{38}$ erg s$^{-1}$ predicted by \citet{Chevalier1989} for the collision of the SN 1987A's ejecta with its circumstellar ring (with $M_1=0.1$, $R_{17}$=5, $\Delta R_{15}=1.6$); but only $L\sim 10^{35}$ erg s$^{-1}$ was actually observed \citep{Burrows2000}, probably due to the drastic difference between the spherical geometry of our models versus the shape of the ring.  As for possible super-luminous X-rays from SN CSM interactions, \citet{Terlevich1992} studied the interaction of supernova with a uniform circumstellar medium of $n\sim 10^7$ cm$^{-3}$ as the basis of a starburst model for active galactic nuclei, and found that the supernova quickly becomes strongly radiative, with most of the X-ray emission coming from the forward shock, which may reach a bolometric luminosity of $10^{43}$ erg s$^{-1}$, consistent with our findings for CSM shells.  

Among the most luminous X-ray SNe ever detected includes SN 2010jl, which was inferred to have an unabsorbed luminosity of $L_X \sim 7\times 10^{41}$ erg s$^{-1}$, most likely from the forward shock front at $\sim 10^{15}$ cm \citep{Chandra2012a}.  However, the actual observed luminosity was initially only 20\% of the unabsorbed luminosity, at least during an early epoch, as the column density was immense: $\sim 10^{24}$ cm$^{-2}$.  Several other SNe have been observed to have X-ray luminosities of a few $10^{41}$ erg s$^{-1}$ more than a year post explosion, for example SN 2008iy \citep{Miller2010} and SN 1995N \citep{Fox2000}.  Indeed, the X-ray light curves of all observed X-ray SNe found in literature have peak luminosities ranging from $10^{37}$ to almost $10^{42}$ erg s$^{-1}$ \citep{Dwarkadas2012}, which may be puzzling given our calculation that $10^{43}$ to $10^{44}$ erg s$^{-1}$ X-ray luminosities with durations of several months are theoretically allowed, albeit contingent on the existence of a CSM shell and some fine-tuning of the shell parameters.  However, almost all of these X-ray SNe were observed below 10 keV, whereas our super-luminous events have enormous CSM shell column densities, and are driven by fast forward shocks reaching temperatures $T\sim 10^9$ K, so before the shock escapes the CSM shell, many unabsorbed X-ray photons from the early emission will have energies $> 10$ keV.  Therefore, the newly launched NuSTAR space telescope, which can observe up to 80 keV, may be better suited for capturing super-luminous X-ray SNe compared with previous satellites; note that the Chandra X-Ray Observatory and Swift's X-Ray Telescope observe below 10 keV, and while the Burst Alert Telescope on Swift can observe up to 150 keV, its poor sensitivity allows it to see $10^{44}$ erg s$^{-1}$ objects only out to $\sim$10 Mpc. 

If the supernova ejecta collides with a dense CSM shell, the shell acts as a wall, resulting in a high reverse shock velocity of $\approx v_{SN} - v_s$.  When the energy initially transmitted into the shell is small, the solutions for the reverse shock have a self-similar nature, and were first solved by \cite{Chevalier1989}.  As the CSM shells in our super-luminous scenarios tend to be much denser than the cavity within (even if the SN ejecta mass is included and averaged over the cavity), it is likely the luminosity of the forward shock running in the CSM shell will dominate the reverse shock retreating into the cavity; furthermore, the reverse shock emission is subject to heavy absorption from the cold, dense shell that condenses between the forward and reverse shocks.

Optical and X-ray emission, interpreted as generated from interactions between SNe ejecta and CS material (or between two ejected shells) have already been used to provide indirect evidence for the explosive ejection of massive CSM shells a few years prior to the supernova, e.g. SN 1994W \citep{Chugai2004,Dessart2009}.  Several observed SNe have a CSM density that seemed to increase with distance from the progenitor:  SN 2008iy \citep{Miller2010}, SN 1996cr \citep{Bauer2008}, SN 2001em \citep{Chugai2006,Schinzel2009}, and SN 2011ja \citep{Chakraborti2013}, suggesting that at least at certain radii, the CSM may be better modeled as a CSM shell rather than a smooth $r^{-2}$ wind.  Also, some of the ultra-luminous X-ray (ULX) sources with luminosities up to $\sim 10^{41}$ erg s$^{-1}$, especially the ones with a thermal spectrum and slow variability, may be due to supernova interacting with massive cirumstellar shells \citep{Swartz2004}.

Aside from converting the kinetic energy of expanding ejecta into radiation upon collision with a massive CSM shell, there is another main mechanism invoked to power super-luminous supernova.  In this mechanism, the SNe explosion launches a shock wave from the center of the star, with the shock heating the material it crosses as the shock travels outward, until the shock escapes at a radius where the material is no longer optically thick to radiation.  More specifically, this \emph{shock breakout} occurs when the photon diffusion timescale becomes shorter than the dynamical timescale of the shock, corresponding to an optical depth of $\tau \simeq c/v_s$ \citep{Weaver1976}; for super-luminous supernova, this edge is at least an order-of-magnitude greater than the edge of the gravitationally-bound progenitor star.  Then, the thermal energy deposited by the shock is gradually emitted as photons diffuse out, analogous to regular Type II-P SNe \citep{Gal-Yam2012}.  

Many ways have been proposed to explain the large effective radii required for superluminous light curves powered via shock breakout, including massive \& optically-thick stellar winds \citep{Ofek2010,Chevalier2011,Moriya2012,Moriya2013}, or massive \& optically-thick shells ejected in prior eruptions \citep{Smith2007,Miller2009} -- assuming that the circumstellar material is optically thick all the way to the CSM shell.  A shock breakout in such environments could also produce X-ray emission \citep{Balberg2011,Katz2011,Chevalier2012,Svirski2012}, and searches for such events have been conducted \citep{Ofek2013a}. The unabsorbed X-ray emission from these shock breakouts can also reach incredible luminosities $\sim 10^{44}$ erg s$^{-1}$, however, the luminosity after shock breakout tends to decline quickly with time, whereas the X-ray emission from optically-thin CSM shell interactions can increase for an extended period of time as the shock runs through and superheats more of the shell.  The collision of SN ejecta with massive CSM shells can also emit much larger total energies in X-rays.  Furthermore, the delay between the optical SN and the X-ray emission is much shorter for shock breakouts.

A non-thermal power-law population of relativistic electrons may be accelerated by the shock.  These could inverse-Compton scatter soft photons and also emit in optical and UV up to X-rays and high energy $\gamma$-rays.  The X-rays from this inverse-Compton component is likely negligible compared to the luminous X-ray emission from the forward shock running through a dense CSM shell considered in this paper, as the former scales with density while the latter scales with density squared; \citet{Chevalier2006} found that during the plateau phase of a Type IIP SN, when the optical flux of the supernova is still $\sim 10^{42}$ erg s$^{-1}$, the inverse Compton X-ray emission is less than $10^{37}$ erg s$^{-1}$, and will further decrease with time as the soft photon flux diminishes.  We should also mention the possibility that the collision of SNe ejecta with massive CSM shells can serve as potential cosmic-ray accelerators \citep{Murase2011}.

\section*{ACKNOWLEDGMENTS}
We thank Sayan Chakraborti, Manos Chatzopoulos, Raffaella Margutti, Takashi Moriya, John Raymond, and Randall Smith for useful discussions.  TP was supported by the Hertz Foundation and the National Science Foundation via a graduate research fellowship.  This work was supported in part by NSF grant AST-0907890 and NASA grants NNX08AL43G and NNA09DB30A.

\bibliographystyle{mn2e}
\bibliography{references}

\end{document}